\documentclass[aps,prb,twocolumn,superscriptaddress]{revtex4}

\usepackage{graphicx}
\usepackage{epstopdf}
\usepackage{hyperref}
\usepackage{amsmath}
\usepackage{amsfonts}
\usepackage{times}
\usepackage{braket}

\usepackage{tikz}

\newcommand{\DM}{\Delta M}
\newcommand{\JO}{J_{\Omega}^{\alpha}}
\newcommand{\JE}{J_{\eta}^{\beta}}
\newcommand{\ri}{\partial_{i}}
\newcommand{\rj}{\partial_{j}}
\newcommand{\figref}[1]{Fig.\ref{#1}}
\newcommand{\zi}{z_{i}}
\newcommand{\zj}{z_{j}}

\begin{document}
\title{Construction of the edge states in fractional quantum Hall
  systems by Jack polynomial}

\author{Ki Hoon Lee}
\affiliation{Asia Pacific Center for Theoretical Physics,
Pohang, Gyeongbuk 790-784, Korea}
\affiliation{Department of Physics, Pohang University of Science and
Technology, Pohang, Gyeongbuk 790-784, Korea}

\author{Zi-Xiang Hu}
\email{zxhu@cqu.edu.cn}
\affiliation{Department of Physics, Chongqing University, 
Chongqing, China}

\author{Xin Wan}
\affiliation{Zhejiang Institute of Modern Physics, Zhejiang
University, Hangzhou 310027, P. R. China}

\date{\today}
\begin{abstract}

We study the edge-mode excitations of a fractional quantum Hall
droplet by expressing the edge state wavefunctions as linear
combinations of Jack polynomials with a negative parameter.  We show
that the exact diagonalization within subspace of Jack polynomials can
be used to generate the chiral edge-mode excitation spectrum in the
Laughlin phase and the Moore-Read phase with realistic Coulomb
interaction. The truncation technique for the edge excitations
simplifies the procedure to extract reliably the edge-mode velocities,
which avoids the otherwise complicated analysis of the full spectrum
that contains both edge and bulk excitations. Generalization to the
Read-Rezayi state is also discussed.

\end{abstract}

\maketitle

\section{Introduction}

The fractional quantum Hall (FQH) effect offers us a unique arena to
study strongly correlated electron systems with a collection of
effective tools, including many-body model
wavefunction,~\cite{laughlin83} exact diagonalization (successful in
often surprisingly small systems), the composite-fermion
theory,~\cite{jainbook} conformal field theories (CFT), and Jack
polynomials (or Jacks).~\cite{bernevig07} Of particular interest
of these bulk-gapped topological phases of matter are gapless edge
excitations, which lead to experimentally measurable
charge~\cite{chang03,marcus08} and neutral currents,~\cite{heiblem10}
including highly nontrivial signals in quasiparticle interference
measurement.~\cite{willett09,willett10,willett12}

Not only are edge excitations key to transport experiments, they also
manifest along artificially cut internal boundary in quantum
entanglement studies, clearly demonstrated in the entanglement
spectrum~\cite{li08} of FQH systems. The multiplicities in the
low-lying part of the entanglement spectrum matches identically with
those of the edge excitations with the corresponding boundary
conditions.  In fact the entanglement spectrum in the case of a
real-space cut~\cite{dubail12a,bernevig12,simon12} can be generated by
a local field theory along the cut.~\cite{dubail12,kitaev06}
 
Explicitly in the disk geometry the model wavefunctions of edge
excitations can be obtained by multiplying the corresponding
ground-state wavefunction by symmetric
polynomials.~\cite{wen1992theory,read96} The Hilbert space of the edge
excitations is robust even in microscopic systems in the presence of
long-range interaction, when their excitation energies are comparable
to those of bulk excitations.~\cite{wan03,wan08} In the CFT construction edge
states can be expressed as the correlators of bulk CFT primary fields
and additional edge fields in the generated chiral
algebra.~\cite{dubail12} In the Laughlin case, the edge-state Hilbert
space can also be generated by a composite-fermion
approach.~\cite{jain10}

Alternatively, Jack polynomials emerged as one of the effective tools
for ground state wavefunction
construction.~\cite{bernevig08,bernevig2008properties,bernevig09}
Jacks are homogeneous symmetric polynomials and Jacks with negative
parameter are shown to be the correlator of the
$\mathcal{W}\text{A}_{k-1}(k+1,k+r)$~\cite{feigin2002differential,feigin2003symmetric,estienne2009relating}
conformal field theory, which includes the $Z_{k}$ parafermion theory.
One advantage of using Jack polynomials is that a recursive
construction algorithm exists, which renders the exact diagonalization
of model Hamiltonians obsolete for a large class of FQH model
wavefunctions.~\cite{bernevig09} The Jack polynomials with a negative
rational Jack parameter $\alpha$ generate ground-state wavefunctions
as well as quasihole wavefunctions,~\cite{yang12} whose constructions
differ only in the corresponding root configurations. By the bulk and
edge correspondance in the FQH systems, this implies that one can use
Jack polynomials to span the space of FQH states on a disk geometry
with single or multiple edge excitations, although they are
generically not orthogonal.

The importance of the inner products of edge states in the CFT
construction has been emphasized in the context of real-space
entanglement spectrum of model FQH states.~\cite{dubail12} In
particular, these inner products take universal values in the
thermodynamic limit, reflecting a correspondance of the edge CFT and
the bulk CFT under generalized screening.

In this paper we present a framework to study edge excitations using
Jack polynomials, whose coefficients are integers and can be
conveniently generated in a computer recursively.~\cite{bernevig09}
The main purpose of this paper is to show that an edge state
wavefunction, expressed as the corresponding ground state wavefunction
(a Jack polynomial itself) multiplied by a symmetric polynomial, can
be written alternatively as a linear combination of several Jacks in
the corresponding momentum subspace. There is a linear map between the
edge-state Hilbert space and the set of admissible root configurations
for Jacks with the proper Jack parameter.  In fact, the coefficients
of the linear combination are universal for all system sizes. By using
Jack polynomial to span the edge-state space, we can facilitate
numerical calculations involving edge states in the presence of
realistic interaction and confinement.  Our paper is organized as
follows. In section~\ref{sec:FQHwf_Jacks} we review general properties
of FQH wavefunctions and basic ingredients of the Jack polynomial
approach of the FQH wavefunctions. We discuss the framework for
constructing the edge Hilbert space by Jack polynomials in
Sec.~\ref{sec:EdgeJacks}. In section~\ref{sec:NumApps} we apply the
approach to generate the edge spectrum for FQH systems in both the
Laughlin and the Moore-Read phases with realistic long-range Coulomb
interaction by exact diagonalization in the edge-excitation
space. Finally, in section~\ref{sec:conclusion}, we summary the paper
and discuss potential applications of the framework. In Appendices, we
provide the details on several crucial statements in the main text on
the Jack polynomial construction of the edge-state Hilbert space.

\section{FQH wavefunctions and Jack polynomials}
\label{sec:FQHwf_Jacks}

\subsection{Basic notations}
The wavefunction of a free particle in the lowest Landau level (LLL)
with the symmetric gauge in a plane is given by
\begin{equation}
\phi _m(z)=\frac{1}{\sqrt{2 \pi 2^mm!}}z^m e^{-z \bar{z} /4},\qquad z=x + i y
\end{equation}
where $m$ is a non-negative integer representing the angular momentum.
Since the Gaussian factor is the same for all $m$, we neglect it in
later discussions and only pay attention to polynomials of $z$.  It is
obvious from the single-particle wavefunction that a many-particle
wavefunction must be a multivariate complex polynomial. Depending on
the statistics of the particles, the polynomial for the corresponding
interacting system is either antisymmetric (for fermions) or symmetric
(for bosons).  Since a bosonic wavefunction can be mapped to a
fermionic wavefunction by the multiplication of a Vandermonde
determinant, the discussion hereafter on bosonic wavefunctions (unless
we specify otherwise) is sufficient for our purposes.

Since symmetric polynomials form a vector space, one needs to choose a
basis.  For the description of bosonic wavefunctions by symmetric
polynomials the basis of \textit{symmetrized monomial} (from now on
monomial implicitly means symmetrized monomial) is the natural choice;
a monomial is nothing but an unnormalized wavefunction for a free
particle in the LLL. We will denote a monomial by a sequence of
integers, which label the occupied orbitals. For instance,
$\pmb{m}_{\{4,2,2,1\}} \equiv \text{Sym}\left (z_{1}^{4} z_{2}^{2}
z_{3}^{2} z_{4}^{1} \right )$. In the presence of the rotational
symmetry, a many-particle wavefunction has a well-defined total
angular momentum $M$. So we only discuss homogeneous symmetric
polynomials, whose basis monomials have the same total degree or total
angular momentum. It means that the sequence of integers that
represents a symmetrized monomial is actually a {\it partition} of the total
degree (an integer), i.e., a non-increasing sequence of nonzero
positive integers whose sum is the integer degree. In the above
example $\{4,2,2,1\}$ is a partition of $9$; in other words, it
represents a monomial with total angular momentum $9$: one particle in
the $z^4$ orbital, two in the $z^2$ orbital, and the other one in the
$z^1$ orbital. Normally, the number of particles in the $z^0$ orbital
is not specified, but its occupation is not ambiguous once the total
number of particles is fixed so we can assume that the unspecified
particles are in the $z^0$ orbital. We denote the {\it weight} (the
sum of the elements) of a partition $\lambda \equiv
\{\lambda_1,\lambda_2,\cdots,\lambda_n\}$ as $\left|\lambda\right| =
\sum_{i=1}^n \lambda_{i}$, or the total degree of its representative
monomial.

Sometimes, it is more convenient to use the occupation representation
(or \textit{configuration}) to represent a basis. The representations
of partition and configuration are of course equivalent, but, just for
the sake of convenience and clarity in discussion, we use both
representations interchangeably. For instance, the following three
notations are considered to be equivalent: $\ket{N-4,1,2,0,1}$,
$\{4,2,2,1\}$, and $\pmb{m}_{\{4,2,2,1\}}$, where $N \ge 5$ is the
total number of particles. Note that we use the curly bracket for the
partition representation and the ket notation for the configuration
representation, respectively. Note that the sequence of integers in a
configuration specify the corresponding numbers of particles in the
orbitals $z^0, z^1, z^2, z^3, z^4, \cdots$, respectively.

One can order distinct partitions of an integer by the lexicographical
order, but in the context of quantum Hall wavefunctions the {\it
  dominance order} is more convenient. The dominance order is
determined by comparing the partial sums of two partitions. Consider
two partition $\lambda$ and $\mu$. If $\sum _{i=1}^{j} \lambda_{i}\geq
\sum _{i=1}^{j}\mu _i$ for every $j$, $\mu$ is dominated by $\lambda
$; the relation is denoted by $\mu \preceq \lambda$. In particular,
when $\mu$ and $\lambda$ are different, we say $\mu$ is {\it strictly
  dominated} by $\lambda $, or $\mu \prec \lambda$. Dominance is
transitive, i.e., if $\nu \preceq \mu$ and $\mu \preceq \lambda$, then
$\nu \preceq \lambda$. The dominance ordering only renders a partial
order to the whole set of partitions of an integer and admits the
lexicographical ordering (i.e., if $\mu \prec \lambda$, $\lambda$ is
larger than $\mu$ in lexicographical order).

One can define an instructive operation called {\it squeezing} on a
partition, which is closely related to the dominance ordering. A
squeezing on a partition moves one pair of particles in the
angular-momentum space closer $\{\cdots,p, \cdots,q,\cdots\}\to
\{\cdots,p-\delta m ,\cdots,q + \delta m, \cdots\}$ (reordering can be
done afterward to make it a partition). Obviously, the total angular
momentum does not change after squeezing. We call the original
partition a {\it parent} and the result a {\it descendant}; the parent
partition strictly dominates the descendant partition.

Applying all the possible sequence of squeezing to a partition (which
one refers to as a {\it root partition}) generates partitions
dominated by the root partition. All these descendant partitions,
together with the root partition, span a Hilbert space (or a symmetric
polynomial space) with a fixed total angular momentum. Studies show
that the Hilbert spaces of a certain series of root partitions are
closely relate to quantum Hall wavefunctions, which can be expressed
as a single Jackpolynomial or the linear combination of a finite set
of Jack polynomials. This latter will be ellaborated when we introduce
Jack polynomial and, in particular, its application to the edge states
of FQH systems. For now, let us denote the Hilbert space spanned by
all the partitions dominated by a root configuration $\lambda$ (i.e.,
$\lambda$ and all the partitions constructed from it by repeated
squeezing) as $H_{\text{sq}}(\lambda)$.

\subsection{A primer on Jack polynomial}
Jack polynomials are homogeneous symmetric
polynomials specified by a root configuration and a rational
parameter. Jacks satisfy a number of differential
equations~\cite{feigin2002differential} and exhibit clustering
properties~\cite{bernevig2008generalized,feigin2003symmetric}. Explicitly,
a Jack is one of the polynomial solutions of the following
Calogero-Sutherland Hamiltonian
\begin{equation}
H_{CS}^{\alpha}(\{z_i\}) = \sum_{i}\left(\zi\ri\right)^2+\frac{1}{\alpha}\sum_{i<j}
\frac{\zi+\zj}{\zi -\zj}\left(\zi \ri -\zj \rj \right) 
\label{CSH}
\end{equation}
where $\alpha$ is a rational parameter and $\partial _{i} \equiv
\partial \left/\partial z_i\right.$.  The definition of a Jack also
requires a root configuration (or partition), such that the Jack
polynomial of a given root configuration $\lambda$ is defined in the
Hilbert space $H_{\text{sq}}(\lambda)$, which is spanned by all the
partitions that can be squeezed from $\lambda$. This fact is due to
the structure of the Calogero-Sutherland Hamiltonian, which only
couples two partitions if one can be squeezed from another. For a
given $\alpha$ we can have a number of Jacks by choosing different
root configuration. If $\alpha$ is positive we can choose any kind of
partition as root partition and these Jacks are all linearly
independent. Therefore, for a positive $\alpha$ Jacks form a basis for
symmetric polynomial.

A recursive construction algorithm exists for Jacks of a root
configuration $\lambda$ and a Jack parameter $\alpha$
\cite{dumitriu2007mops,thomale10decompose}.  Hence one can compute the
coefficients of the following monomial expansion of a Jack
\begin{equation}
J_{\lambda}^{\alpha}(\{z\}) =\pmb{m}_{\lambda}+ 
\sum _{\mu \prec \lambda}c_{\mu}\pmb{m}_{\mu}
\end{equation}
The convention is to fix the coefficient of the root configuration as
unity and other coefficients are scaled accordingly. The use of this
set of coefficients in subsequent numerical calculations for a given
geometry (e.g., disk or sphere) often requires a proper normalization
for each monomial.

Some Jacks of a negative $\alpha$ can be directly related to FQH
wavefunctions. It was conjectured that Jacks with a negative parameter
with proper root configuration are correlators of certain conformal
theories~\cite{feigin2002differential,feigin2003symmetric}. Bernevig
and Haldane and co-workers further explored the idea extensively and
showed that the Jack polynomial approach is an efficient and
insightful development to obtain and to exploit FQH
wavefunctions~\cite{bernevig08,bernevig2008properties,bernevig2008generalized,bernevig2009clustering,bernevig09}.
It was proven for general $(k,r)$ that Jacks are the correlators of
$\mathcal{W}\text{A}_{k-1}(k+1,k+r)$ conformal
theories~\cite{bernevig2009central,estienne2009relating}.  In other
words, the correlators are eigenfunctions of the Calogero-Sutherland
Hamiltonian [Eq.~\eqref{CSH}]. Further more, the application of Jack
polynomial on a certain type of quasihole wavefunctions was also
proposed and proven to be correct with the discovery of interesting
duality structure between electron and quasihole
wavefunctions\cite{bernevig08,estienne2010electron}.

In contrast to a positive $\alpha$, we have, for a negative $\alpha$,
restrictions on the choice of root configuration. One allowed choice
is to use a $(k,r)$ admissible root configuration, which leads to
legitimate FQH trial wavefunctions. The $(k,r)$-admissibility means
that there can be at most $k$ particles in $r$ consecutive
orbitals. More precisely, a partition $\lambda$ is said to be $(k,r)$
admissible, if $\lambda_{i}-\lambda_{i+k}\geq r$. The densest $(k,r)$
admissible root configuration and a corresponding Jack parameter
$\alpha = -(k+1)/(r-1)$ (with the condition that $k+1$ and $r-1$ are
coprime) generate the FQH ground state with a filling fraction $\nu
=k/r$ (corresponding to $\nu =k/(k+r)$ in the fermionic case). Note
that $\alpha$ is negative.

\begin{table}
\begin{tabular}{|c|l|c|l|}
\hline
$\left(k,r\right)$  & bosonic & $\left(k,k+r\right)$ &fermionic  \\
\hline
$\left(1,2\right)$ & $\ket{1,0,1,0,1,\cdots,1}$ & $(1,3)$ &$\ket{1,0,0,1,0,0,1,\cdots ,1}$\\
$\left(1,4\right)$ & $\ket{1,0,0,0,1,\cdots,1}$ & $(1,5)$ &$\ket{1,0,0,0,0,1,\cdots,1}$\\
$\left(2,2\right)$ & $\ket{2,0,2,0,2,\cdots,2}$ & $(2,4)$ &$\ket{1,1,0,0,1,1,\cdots ,1,1}$\\
\hline	
\end{tabular}
\caption{The densest $(k,r)$ admissible configurations. Fermionic
  configuration is $(k,k+r)$ admissible configurations which is the
  counter part state of $(k,r)$ admissible bosonic configuration.}
\label{groundpartitiontable}
\end{table}

The applicability of the Jacks to the fermionic case is highly
nontrivial since the multiplication of a Jack (with a monomial
expansion) with a Vandermonde determinant cannot be straightforwardly
mapped to a sum of Slater determinants. Nevertheless, a recursive
procedure was found to generate the Slater-determinant expansion of
$\prod_{i<j}\left(z_i-z_j\right) J_{\lambda}^{\alpha}(\{z\})$
\cite{bernevig09,thomale10decompose}. Hence, all the discussion on
bosonic wavefunctions, including numerical calculation, is applicable
to fermionic wavefunctions in practice.

\section{Jack polynomials for FQH edge states}
\label{sec:EdgeJacks}

\subsection{Laughlin edge states: an example}

In the simplest case of an Abelian Laughlin state, edge excitations
are deformation of edge density of the incompressible liquid. One can
write down the trial wavefunction for an edge state as a symmetric
polynomial multiplied by the ground state
wavefunction~\cite{wen1992theory}
\begin{equation}
\Psi (\{z\}) = S(\{z\}) \Psi _L^{r}(\{z\}) 
\label{LaEdge}
\end{equation}
where $\Psi _{L}^{r}=\prod _{i<j}^n\left(z_i-z_j\right)^r$ is the
Laughlin state at a filling fraction $\nu = 1/r$ (with total angular
momentum of $M_0 = rN(N-1)/2$) and $S(\{z\})$ a homogeneous symmetric
polynomial. In this expression we encounter the multiplication of a
monomial by another, which is then to be expanded as a sum of
monomials. The expansion is not straightforward and often
computationally expensive. Similarly, a symmetrized monomial
multiplied by a antisymmetric Vandermonde determinant is in general a
complex superposition of antisymmetrized monomials (or Slator
determinants). The same difficulty also arises in the wavefunctions of
quasihole states. Therefore, the nontrivial multiplication of
physically natural basis states hinders us to efficiently use ansatz
edge-state or quasihole-state wavefunctions in computation. The
difficulty, however, can be resolved by the introduction of Jack
polynomial~\cite{bernevig09}. In this paper, we focus on the
applications for the edge-state wavefunctions.

The edge state, as well as the ground state, is a zero-energy
eigenstate of an ideal Hamiltonian with short-range interaction. One
remarkable fact about Eq.~\eqref{LaEdge} is that the number of
linearly independent edge states for a given momentum ($M = M_0 +
\Delta M$) predicted by the chiral Luttinger liquid theory is the same
as the number of the symmetric polynomials of degree $\Delta M$. To
see this, it is sufficient to check the dimension of the legitimate
symmetric polynomials for the given momentum $M$. By counting the
number of possible monomials of degree $\Delta M$ or possible
partitions of $\Delta M$, we obtain the dimension of the edge-state
subspace with a total angular momentum $M = M_0 + \Delta M$. Hence the
subspace dimension for the given momentum $\DM$ is exactly the number
of partition of the integer $\DM$ and the counting is consistent with
the edge theory of a single branch of chiral bosons.

Given the fact that the Laughlin state can be expressed as a single
Jack with a root configuration being a $(k,r)$ admissible partition of
$M_0$, one natually wants to explore the connection of the edge states
to Jacks with root configurations being $(k,r)$ admissible partitions
of $M > M_0$. The clustering property is enforced such that the Jacks
are also zero-energy states of the ideal Hamiltonian. In other words,
these Jacks are edge states. One immediate question is whether the
distinct Jacks span the same edge-state Hilbert space. A harder
question is how to relate the edge states, given by
Eq.~\eqref{LaEdge}, and the Jacks of various root configurations; they
are in general not orthorgonal to one another within each group.
Because one can form a quasihole wavefunction by a linear combination
of edge excitations with proper weight, the construction of edge
states by superposing Jack polynomials leads straightforwardly to the
construction of quasihole
wavefunctions~\cite{bernevig08,estienne2010electron} in the same
manner. 

\subsection{Edge Jacks polynomials}

As we argued above, admissible root configurations with larger angular
momentum (than the ground state anngular momentum) lead to Jacks in
the Hilbert space of edge states because of the clustering property of
the Jacks. But so far we do not know the explicit relation between
these Jacks and the ansatz wavefunctions for edge states, which is
crucial if one wants to efficiently expand any edge-state wavefunction
into monomials.

To explicitly show that the Jacks with $(k,r)$ admissible partitions
span the same Hilbert space as the edge states described by
Eq.~\eqref{LaEdge}, we introduce a compact notation for the admissible
root configurations.  In this notation the admissible root
configuration for an edge state is obtained by adding to the ground
state root configuration $\Omega$ a partition $\eta$ of $\Delta M$, or
$\lambda = \Omega+\eta$.  Note that the total angular momentum of the
Jacks wavefunction of $\lambda$ is $\left|\Omega\right|
+\left|\eta\right|$, or $M = M_0 + \Delta M$.  Hence, once $(k,r,N)$
is fixed one can label a state only by $\eta$, which we refer later as
the {\it edge partition}, without any ambiguities. As a concrete
example, we consider the Laughlin sequence, for which $k=1$ and $r$ is
a certain positive even integer. The $(1,r)$ admissibility of the
densest ground state means $\Omega_{i} - \Omega_{i+1} = r$. The
difference changes to $\lambda _{i+1}-\lambda _{i} = (\Omega_{i} -
\Omega_{i+1}) + (\eta _{i}-\eta _{i+1}) = r + (\eta _{i}-\eta
_{i+1})$, when $\eta$ is added to $\Omega$. Therefore, the $(1,r)$
admissible condition of $\lambda = \Omega + \eta$ is that $\eta$ is a
nondecreasing sequence of integers, which precisely means that $\eta$
is a partition of the additional angular momentum $|\eta| = \Delta
M$. This identification of $\eta$ gives exactly the expected number of
the edge states described by Eq.~\eqref{LaEdge}.  Note that $\eta$ is
meaningful for any system with $N \ge \Delta M$ and, in this case, the
number of $(1,r)$ admissible root configurations, hence of linearly
independent edge states, are independent of the particle number $N$.
The analysis of $(k,r)$ admissible states with $k > 1$ needs more
care, as $\eta$ cannot be a simple partition. We defer the discussion
to Appendix~\ref{app:counting}.

These admissible root partitions themselves are related by squeezing
or dominance ordering. Regardless of $(k,r)$, the greatest partition
for a given $\DM$ is $\lambda_{\DM} = \Omega + \{\DM\}$ and the other
admissible partitions are dominated by it. This implies that the
Hilbert space for a Jack with an admissible root partitions is a
subspace of $H_{sq}(\lambda_{\DM})$. Moreover, $H_{sq}(\lambda_1)
\subset H_{sq}(\lambda_2)$ if the two partitions satisfy $\lambda_1
\prec \lambda_2$ by dominance ordering. It means that the dominance
relation between the root configurations is sufficient to describe the
inclusion relation of the Hilbert space for the Jacks generated by the
respective root configurations.

Now we consider the concrete construction of the edge-state space for
the Laughlin case, i.e. $k=1$.  The dominance relation between
$\lambda$s is same as the corresponding $\eta$s.  Therefore, in lieu
of Eq.~\eqref{LaEdge} one may expect a set of equations that relate
the Jacks with the edge partition $\eta$s as their roots and the Jacks
with the full partition $\lambda$s. Indeed, we find
\begin{equation}
J_{\Omega+\eta}^{\alpha}=\JE \JO
\label{LAEDGEEQ}
\end{equation} 
where $\alpha=-\frac{2}{r-1}$ and $\beta= \frac{2}{r+1}$. We leave the
derivation in Appendix~\ref{app:LaEdgeEq}. Note that there is no
restriction on the root configuration for the positive rational
parameter $\beta$. The result means that $\JE$ is a representation of
$J_{\Omega+\eta}^{\alpha}$ that can be readily used to decompose any
edge state in the form of Eq.~\eqref{LaEdge}. For an arbitrary
symmetric polynomial $S({z})$, one can expand it in terms of Jacks
$\JE$, since the Jacks form a basis for the homogeneous symmetric
polynomial space. Therefore, it is possible to obtain the monomial
expansion of $S({z}) \JO $ efficiently.

As an important example, we consider the monomial expansion of an
arbitrary monomial multiplied by the Laughlin ground state
$J_{\Omega}^{\alpha}$ for $\alpha=-2/(r-1)$. In other words, we want
to calculate the coefficient $c_{\mu}$ in the expression
\begin{equation}
\pmb{m}_{\eta}J_{\Omega}^{\alpha} = \sum _{\mu \preceq
  \Omega+\eta}c_{\mu}\pmb{m}_{\mu}.
\end{equation} 
This can be easily achieved once we know the monomial expansion of
$\JE$. By inverting the relation between $\pmb{m}_{\eta}$s and $\JE$s,
one can express any $\pmb{m}_{\eta}$ as a linear combination of
$\JE$s.  In other words, one then obtains a Jack-polynomial expansion
of the monomial multiplied by the ground state wavefunction. We can,
equivalently, write down the array of equations in a compact form for
any $\DM$ as
\begin{equation}
\label{eq:Rel}
\mathbf{J} \mathbf{m} \JO = \mathbf{j}_{E}, 
\end{equation}
where $\mathbf{j}_{E}$ is a column vector whose components are the
edge Jacks $J^{\alpha}_{\Omega+\eta}$ and $\mathbf{m}$ a vector of
$\pmb{m}_{\eta}$s; both are ordered in the lexicographical order of
$\eta$. Explicitly, we have
\begin{eqnarray}
\mathbf{j}_{E} &=&
\left(J^{\alpha}_{\Omega+\{1,\cdots,1\}},\cdots,J^{\alpha}_{\Omega+\{\DM\}}
\right)^{T},\\ \mathbf{m} &= &
\left(\mathbf{m}_{\{1,\cdots,1\}},\cdots,\mathbf{m}_{\{\DM\}}\right)^{T}.
\end{eqnarray}
They have the same dimension, which is the number of all possible
partitions of $\DM$.  $\mathbf{J}$ is a matrix, which consists of the
corresponding coefficients of $\JE$ in terms of $\pmb{m}_{\nu}$, i.e.,
\begin{equation}
\mathbf{J}_{\eta \nu} = c^{(\eta)}_{\nu},
\end{equation}
if we assume
\begin{equation}
J^{\beta}_{\eta} = \sum_{\nu \preceq \eta} c^{(\eta)}_{\nu} \mathbf{m}_{\nu}.
\end{equation}
Again, we recall the convention $c^{(\eta)}_{\eta} = 1$. By inverting 
$\mathbf{J}$, we obtain 
\begin{equation}
\label{eq:invRel}
\mathbf{m} J_{\Omega}^{\alpha} = \mathbf{J}^{-1} \mathbf{j}_{E}.
\end{equation}
Since the rank of $\mathbf{J}$ is only the number of distinct
partitions of $\DM$, the computation of $\mathbf{J}$ and its inverse
is computationally cheap. 

We point out two important properties of $\mathbf{J}$ in the
following. Firstly, when we arrange the basis (for both the Jacks by
their root partitions and the monomials) in lexicographical order or
any other order that respects the dominance ordering, we obtain a
lower triangular matrix with all identities at its diagonal (i.e., a
unitriangular matrix) for $\mathbf{J}$.  In other words, if $\mu >
\lambda$ by dominance ordering or has no relation between $\mu
J_{\Omega}^{\alpha}$ and $\lambda$, then the inner product of
$\mathbf{m}_{\mu}$ and $J_{\lambda}$ is zero, as the two share no
common monomials in their respective expansions. We note that
$H_{sq}(\Omega)H_{sq}(\eta) \subset H_{sq}(\Omega + \eta)$ for our
understanding, which can be proved as in Appendix~\ref{app:proof} and
which indicates that $\mathbf{m}_{\eta} J_{\Omega}^{\alpha} \subset
H_{sq} (\Omega + \eta)$.  Secondly, as long as the number of particles
is larger than $\DM$, $\mathbf{J}$ is independent of system size. As
an example, let us define $\eta_{1} = \{1,1,1,1\}$, $\eta_{2} =
\{2,1,1\}$, $\eta_{3} = \{2,2\}$, $\eta_{4} = \{3,1\}$, and $\eta_{5}
= \{4\}$ for $k = 1$, $r = 2$, and $\DM = 4$.  We have
\begin{equation}
\left (
\begin{array}{c}
\mathbf{m}_{\eta_{1}} \\
\mathbf{m}_{\eta_{2}} \\
\mathbf{m}_{\eta_{3}}  \\
\mathbf{m}_{\eta_{4}} \\
\mathbf{m}_{\eta_{5}}
\end{array} 
\right )
\JO
 = 
 \left(
 \begin{array}{ccccc}
  1 & 0 & 0 & 0 & 0 \\
  -\frac{36}{11} & 1 & 0 & 0 & 0 \\
  \frac{27}{22} & -\frac{6}{5} & 1 & 0 & 0 \\
  \frac{27}{11} & -\frac{27}{25} & -\frac{6}{5} & 1 & 0 \\
  -\frac{81}{55} & \frac{36}{25} & \frac{6}{35} & -\frac{4}{3} & 1
 \end{array}
 \right)
\left ( 
\begin{array}{c}
J^{\alpha}_{\Omega+\eta_{1}} \\
J^{\alpha}_{\Omega+\eta_{2}} \\
J^{\alpha}_{\Omega+\eta_{3}} \\
J^{\alpha}_{\Omega+\eta_{4}} \\
J^{\alpha}_{\Omega+\eta_{5}} 
\end{array} 
\right )
\end{equation} 
where the matrix on right hand side is $\mathbf{J}^{-1}$, the inverse of 
\begin{equation}
\mathbf{J} =
\left(
\begin{array}{ccccc}
 1 & 0 & 0 & 0 & 0 \\
 \frac{36}{11} & 1 & 0 & 0 & 0 \\
 \frac{27}{10} & \frac{6}{5} & 1 & 0 & 0 \\
 \frac{108}{25} & \frac{63}{25} & \frac{6}{5} & 1 & 0 \\
 \frac{72}{35} & \frac{12}{7} & \frac{10}{7} & \frac{4}{3} & 1
\end{array}
\right)
\end{equation}
The electron number independence of $\mathbf{J}$ is guaranteed by
product rule \cite{bernevig09,thomale10decompose} of Jacks. If the
electron number $N$ is larger than $\DM$, then the admissible root
partitions are in one-to-one correspondance with those of the system
with $N = \DM$. Hence by the product rule, the coefficients of the
admissible root partitions remain unchanged with increasing system
size.

Before we move on to the more general cases (i.e., $k > 1$), we
emphasize that the form of $\mathbf{J}$ for $k = 1$ is the direct
consequence of Eq.~\eqref{LAEDGEEQ}. We do not know any such simple
relation for $k > 1$, hence the form of $\mathbf{J}$ is not as simple
but still can be computed. The reason for the complication is because
an edge sequence is no longer a simple partition, but consists of $k$
subpartitions. However, the Jacks with admissible root configurations
still exhaust the Hilbert space of the edge states, therefore, we can
still expand any edge-state wavefunction as a linear combinations of
the edge Jack polynomials. That is to say, we look for a
generalization of Eq.~(\ref{eq:Rel}) or Eq.~(\ref{eq:invRel}). As the
edge space is no longer spanned by simple monomials multiplied by the
ground state. Nevertheless, we can still write down a complete set of
edge states for each $\Delta M$ (which we label as $\mathbf m_E$ that
replaces ${\mathbf m} \JO$), as Milovanovi\'c and Read did for the $k
= 2$ Moore-Read case. Now the question converts to how to create a
dictionary $\mathbf J$ to translate $\mathbf m_E$, which is normally
easier for analytical discussions, to the set of edge Jack polynomials
$\mathbf j_E$, which is more convenient for numerical calculations.
Note that, in the basis of monomials of degree $M_0 + \Delta M$
[i.e. in the Hilbert space $H_{sq}(\Omega+\{\DM\})$], $\mathbf{m}_E$
and $\mathbf{j}_E$ can be viewed, alternatively, as matrices. A key
observation is that the dimension of edge state (number of rows) is
usually much smaller than the dimension of $H_{sq}(\Omega+\{\DM\})$
(number of columns), such that we do not need to know all elements
of $\mathbf{m}_E$ to obtain $\mathbf{J}$; in other words, we have a
set of overdetermined linear equations. The most obvious solution to
the problem which subspace in $H_{sq}(\Omega+\{\DM\})$, whose
dimension is the same as that of the edge space, should be chosen to
determine $\mathbf{J}$ turns out to be the space spanned by admissible
root partitions.

This can be further extended to express, in terms of Jacks, quasihole
states (including multiple quasiholes), which can be written as the
superpositions of edge states with powers of the quasihole coordinates
as coefficient and which we will not elaborate.

\section{Numerical applications}
\label{sec:NumApps}

\subsection{Diagonalization in the truncated space of edge states}

One of the applications of the edge Jacks is the exact diagonalization
of the Hamiltonian with realistic Coulomb interaction within the
edge-state subspace spanned by the Jacks. The exact diagonalization in
the full Hilbert space is difficult bacause of its exponentially
growing size as particle number increases. Fortunately, the low-energy
eigenfunctions of the Hamiltonian are well represented by CFT
correlators or Jacks so we can overcome this difficulty by projecting
the Hamiltonian to the edge-state subspace.  For electrons we use the
fermionic Jacks, which are symmetric Jacks multiplied by a Vandermonde
determinant. The recursive construction procedure is well documented
in Ref.\onlinecite{bernevig09}. Here we consider a realistic
Hamiltonian
\begin{equation}
H = \frac{1}{2} \sum_{mnl} V^{l}_{mn} c_{m+l}^{\dagger} c_{n}^{\dagger} c_{n+l} c_{m} 
+ \sum_{m} U_{m} c_{m}^{\dagger} c_{m},
\end{equation}
where the Coulomb matrix elements $V^{l}_{mn}$ are
\begin{equation}
   V^{l}_{mn}=\int d^2 r_1\int d^2 r_2 \phi _{m+l}^*
 (\vec{r}_1 )\phi_{n}^* (\vec{r}_2)
      \frac{{e^2 }}{{\varepsilon r_{12} }}
 \phi _{n+l} (\vec{r}_1)\phi _{m} (\vec{r}_2),
 \end{equation}
and $U_m$ is the matrix element of the rotationally invariant
confining potential due to the disk-shaped positive background charge
at a setback distance $d$
\begin{equation}
  U_m = {N_e e^2 \over \pi R^2 \varepsilon } \int d^2 r \int_{\rho
  < R} d^2 \rho \frac{|\phi _m (\vec{r})|^2}{\sqrt{|\vec{r} -
 \vec{\rho}|^2 + d^2}}. 
\end{equation} 
Here $\varepsilon$ is the dielectric constant and $R$ 
the radius of the background charge disc. 

\begin{figure}[h]
\includegraphics[width=240px]{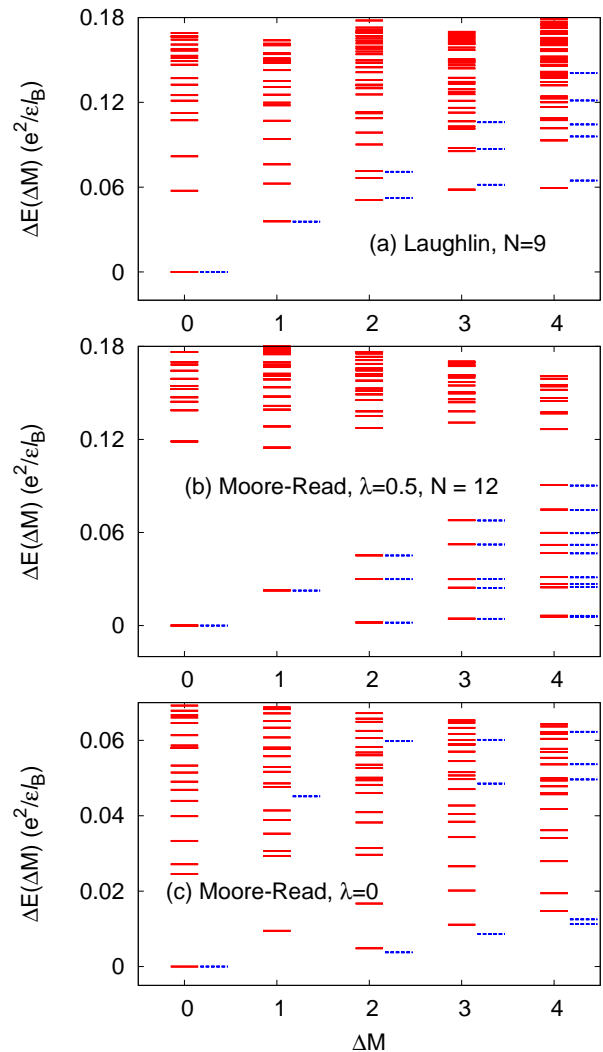}
\caption{The comparison of the low-energy excitation spectra from the
  exact diagonalization in the full Hilbert space (red bars) and from
  the diagonalization within the edge Jack polynomial subspace (blue
  bars with an artificial shift in momentum for clarity). (a) The
  Laughlin case with 9 electrons. (b) and (c) are for Moore-Read case
  with 12 electrons.  Here, $\lambda$ is the parameter for the mixed
  Hamiltonian with three-body interaction $H = (1-\lambda)
  H_{\text{Coulomb}} + \lambda H_{3B}$, as in
  Ref.~\onlinecite{wan08}. $\lambda = 0$ is the case of pure Coulomb
  interaction.}
\label{fig:N9}
\end{figure}

The diagonalization can be performed in the subspace of Jacks with a
fixed momentum due to the rotational invariance. For instance, for
$\DM=3$, available $\eta$s of Laughlin sequence are $\{3\}$,$\{2,1\}$
and $\{1,1,1\}$. Hence the subspace we deal with are only $3$
dimension consists in $J_{\Omega+\{3\}},J_{\Omega+\{2,1\}}$ and
$J_{\Omega+\{1,1,1\}}$.  Since Jacks are not orthogonal to each other,
orthogonalization is needed. We project the Hamiltonian to the
subspace spanned by the orthogonalized edge basis and then perform the
exact diagonalization. The result for 9 electrons at $d=0.6 l_B$ is
shown in \figref{fig:N9}(a).  The low-lying spectrum is in good
agreement with that from the exact diagonalization in the full Hilbert
space. A similar comparison for the Moore-Read phase with 12 electrons
are shown in \figref{fig:N9}(b) and (c). We illustrate here, as in
Ref. \onlinecite{wan08}, the diagonalization results for a mixed
Hamiltonian with both Coulomb and three-body interaction, i.e., $H =
(1-\lambda) H_{\text{Coulomb}} + \lambda H_{3B}$.  Explicitly, the
three-body interaction $H_{3B}$, which generates the Moore-Read
wavefunction as its exact ground state, has the form
\begin{equation}
\label{eqn:threebody} H_{3B} = -\sum_{i < j <
k}S_{ijk}[\nabla^2_i\nabla^2_j (\nabla^2_i+\nabla^2_j)\delta({\bf r}_i -
{\bf r}_j)
\delta({\bf r}_i -{\bf r}_k)],
\end{equation}
where $S$ is a symmetrizer:
$S_{123}[f_{123}]=f_{123}+f_{231}+f_{312}$, where $f$ is symmetric in
its first two indices. When $\lambda = 0.5$, the edge states, which
are zero-energy states for $H_{3B}$, pick up finite energies but are
still well separated from the bulk states in small excitation-momentum
sectors. \figref{fig:N9}(b) shows that the edge-state spectrum agrees
well with that from the full diagonalization. In the case of the pure
Coulomb interaction in \figref{fig:N9}(c), the agreement is not as
good due to the mixing of the edge states with the bulk
states.~\cite{wan08}

From the comparison of the excitation spectra in \figref{fig:N9} of
the full-space diagonalization and the truncated-space diagonalization
we have the following observation. For small excitation-momentum
sectors, the energies are almost identical in two cases.  Even for
$\nu=5/2$ state with the Coulomb interaction, their difference
decreases as the system size increases and vanishes in the
thermodynamic limit. This is particularly useful to extract the edge
velocities, to be discussed in the next subsection, from the
truncated-space diagonalization for system sizes significantly larger
than those can be handled by the full diagonalization. For regular
computing systems, the main bottleneck of this method for large
systems is the large storage space (or memory) for the edge Jacks.

\subsection{Edge-Mode Velocities}

Edge-mode velocities in the Moore-Read phase and the Read-Reazyi phase
are important non-universal quantities, as they are closely related to
the decoherence length of the non-Abelian quasiparticles propagating
along the edge of, say, a quasiparticle interferometer. Their
calculation can be made more efficient with the help of the
diagonalization in the space spanned by the edge Jacks.  Charge mode
velocity can be defined as $v_{b} = L(E_{0}(\DM=1) - E_{0}(\DM=0))
[e^{2}/\epsilon \hbar] $, where the perimeter of the quantum droplet
is $L = 2\pi R = \sqrt{2N(k+r)/k}$. Neutral mode velocity can be
defined as $v_{f} = L(E_{0}(\DM=2) - E_{0}(\DM=0))/2 [e^{2}/\epsilon
  \hbar]$.~\cite{hu2009edge} $E_{0}(\DM)$ is the lowest eigenenergy
for the given momentum $\DM$. Through finite-size scaling we can
extrapolate the edge-mode velocities in the thermodynamic limit. We
have demonstrated the applicability of the finite-size scaling with
data from the diagonalization in the full Hilbert space in
Ref.~\onlinecite{hu2009edge}. 

\begin{figure}[h]
\includegraphics[width=240px]{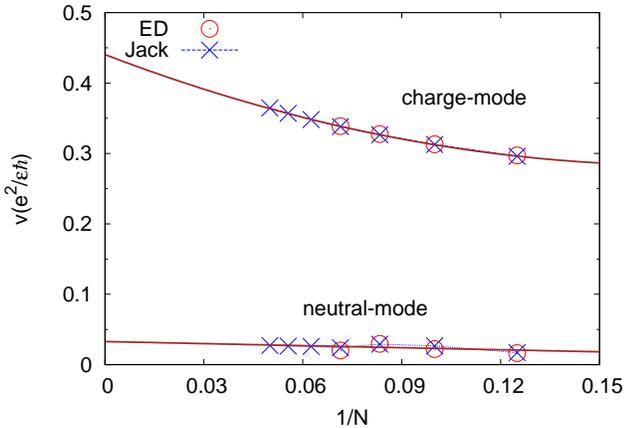}
\caption{Finite-size scaling of the edge-mode velocities for $\nu =
  5/2$ with Coulomb interaction and confinement from neutralizing
  background charge.  The charge-mode velocity
  is fitted by a quadratic function in $1/N$ and, in the thermodynamic
  limit, the velocity reads $0.44 e^{2}/\epsilon \hbar $ (based on
  data from 8-20 particles). The neutral-mode velocity is fitted by a
  linear dependence on $1/N$. In the thermodynamic limit, the velocity
  reads $0.033 e^{2}/\epsilon \hbar$.}
\label{fig:mrvelocity}
\end{figure}

We first apply the velocity calculation to the 1/2-filling Moore-Read
case, i.e. $k=2$ and $r=2$, in the first excited Landau level. With
the diagonalization in the edge-state space we can handle systems with
up to 20 electrons. Fig.~\ref{fig:mrvelocity} shows that a charge-mode
velocity of $0.44 e^{2}/\epsilon \hbar $ can be extrapolated in the
thermodynamic limit.  This value agrees with the previous finite-size
scaling result based on the exact diagonalization of systems of 8-14
electrons.~\cite{hu2009edge}.  The neutral-mode velocity suffers more
from the the finite-size effect due to its smallness and can be
extrapolated to be $0.033 e^{2}/\epsilon \hbar$ in the thermodynamic
limit. We confirm the sharp contrast in the magnitude of the
charge-mode and neutral-mode velocities. As discussed in
earlier works,~\cite{wan08,hu2009edge} this sets an upper limit of
about 1 $\mu$m for the coherence length for charge-e/4 quasiparticles
at 5/2 filling factor in a Fabry-Perot interferometer. Recently,
Willett and co-workers~\cite{willett13} reported such coherence length
to be 0.49-0.74 $\mu$m, consistent with our study of the edge-mode
velocities in the realistic model.

We also attempted the same analysis of edge velocities for the
Read-Rezayi state at filling fraction 13/5 with Coulomb interaction,
whose particle-hole conjugated state at 12/5 filling has been observed
in some experiments~\cite{xia04,kumar10,Zhangchi2012}. Preliminary results suggest that, again for
$d/l_B = 0.6$, the charge-mode velocity is $0.537 e^{2}/\epsilon
\hbar$ (roughly 6/5 times the charge-mode velocity at 5/2 filling as
expected) while the neutral-mode velocity is extrapolated to almost
zero (more precisely, another order of magnitude smaller than the
neutral-mode velocity in the Moore-Read case). The smallness of the
neutral-mode velocity is a strong indication that the Read-Rezayi is
very fragile in reality. The small value can be very sensitive to
Landau level mixing, which remains to be explored.  Nevertheless, we
expect that the qualitative conclusion that the neutral-mode velocity
is much smaller than the charge-mode velocity will still be valid.

\section{Conclusion} 
\label{sec:conclusion}

In summary, we discuss the identification of the Hilbert space of edge
excitations in the Laughlin--Moore-Read--Read-Rezayi series as spanned
by the appropriate Jack polynomials with admissible root
configurations. In particular, we elaborate on the Laughlin case ($k =
1$), which contains a single charged mode. We explain how to establish
a linear map between the polynomial edge wavefunctions and edge Jack
polynomials. The map is of particular use in the presence of realistic
interaction and confinement, in which case the edge spectrum has a
nontrivial dispersion with a nonuniversal edge-mode velocity. The
mapping formalism and its numerical application can be generalized to
larger $k$s. The applicability has been checked by the comparison of
the edge spectra obtained in the truncated edge Jack polynomial space
and in the full Hilbert space with exact diagonalization. The
truncation approach has the advantage of being able to treat larger
systems than the conventional exact diagonalization approach. As an
example, we are able to calculate the edge-mode velocities in the
Moore-Read phase for $\nu = 5/2$ up to 20 electrons and confirm with
greater confidence that the neutral-mode velocity is an order of
magnitude smaller than the charge-mode velocity.

This work was supported by the 973 Program under Project
No. 2012CB927404 and NSFC Projects No. 11174246 and
No. 11274403. K.H.L. acknowledges the support at the Asia Pacific
Center for Theoretical Physics from the Max Planck Society and the
Korea Ministry of Education, Science and Technology.

\appendix

\section{Edge partitions for $k > 1$}
\label{app:counting}
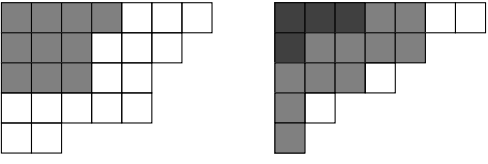
\begin{figure}
\parbox{1.4in}{
\begin{tikzpicture}
\filldraw[fill=gray,draw=black] (0.,-0.4) rectangle (0.4,0.);
\filldraw[fill=gray,draw=black] (0.4,-0.4) rectangle (0.8,0.);
\filldraw[fill=gray,draw=black] (0.8,-0.4) rectangle (1.2,0.);
\filldraw[fill=gray,draw=black] (1.2,-0.4) rectangle (1.6,0.);
\draw (1.6,-0.4) rectangle (2.,0.);
\draw (2.,-0.4) rectangle (2.4,0.);
\draw (2.4,-0.4) rectangle (2.8,0.);
\filldraw[fill=gray,draw=black] (0.,-0.8) rectangle (0.4,-0.4);
\filldraw[fill=gray,draw=black] (0.4,-0.8) rectangle (0.8,-0.4);
\filldraw[fill=gray,draw=black] (0.8,-0.8) rectangle (1.2,-0.4);
\draw (1.2,-0.8) rectangle (1.6,-0.4);
\draw (1.6,-0.8) rectangle (2.,-0.4);
\draw (2.,-0.8) rectangle (2.4,-0.4);
\filldraw[fill=gray,draw=black] (0.,-1.2) rectangle (0.4,-0.8);
\filldraw[fill=gray,draw=black] (0.4,-1.2) rectangle (0.8,-0.8);
\filldraw[fill=gray,draw=black] (0.8,-1.2) rectangle (1.2,-0.8);
\draw (1.2,-1.2) rectangle (1.6,-0.8);
\draw (1.6,-1.2) rectangle (2.,-0.8);
\draw (0.,-1.6) rectangle (0.4,-1.2);
\draw (0.4,-1.6) rectangle (0.8,-1.2);
\draw (0.8,-1.6) rectangle (1.2,-1.2);
\draw (1.2,-1.6) rectangle (1.6,-1.2);
\draw (1.6,-1.6) rectangle (2.,-1.2);
\draw (0.,-2.) rectangle (0.4,-1.6);
\draw (0.4,-2.) rectangle (0.8,-1.6);
\end{tikzpicture}
}
\parbox{1.4in}{
\begin{tikzpicture}
\filldraw[fill=darkgray,draw=black] (0.,-0.4) rectangle (0.4,0.);
\filldraw[fill=darkgray,draw=black] (0.4,-0.4) rectangle (0.8,0.);
\filldraw[fill=darkgray,draw=black] (0.8,-0.4) rectangle (1.2,0.);
\filldraw[fill=gray,draw=black] (1.2,-0.4) rectangle (1.6,0.);
\filldraw[fill=gray,draw=black] (1.6,-0.4) rectangle (2.,0.);
\draw (2.,-0.4) rectangle (2.4,0.);
\draw (2.4,-0.4) rectangle (2.8,0.);
\filldraw[fill=darkgray,draw=black] (0.,-0.8) rectangle (0.4,-0.4);
\filldraw[fill=gray,draw=black] (0.4,-0.8) rectangle (0.8,-0.4);
\filldraw[fill=gray,draw=black] (0.8,-0.8) rectangle (1.2,-0.4);
\filldraw[fill=gray,draw=black] (1.2,-0.8) rectangle (1.6,-0.4);
\filldraw[fill=gray,draw=black] (1.6,-0.8) rectangle (2.,-0.4);
\filldraw[fill=gray,draw=black] (0.,-1.2) rectangle (0.4,-0.8);
\filldraw[fill=gray,draw=black] (0.4,-1.2) rectangle (0.8,-0.8);
\filldraw[fill=gray,draw=black] (0.8,-1.2) rectangle (1.2,-0.8);
\draw (1.2,-1.2) rectangle (1.6,-0.8);
\filldraw[fill=gray,draw=black] (0.,-1.6) rectangle (0.4,-1.2);
\draw (0.4,-1.6) rectangle (0.8,-1.2);
\filldraw[fill=gray,draw=black] (0.,-2.) rectangle (0.4,-1.6);
\end{tikzpicture}
}
\caption{\label{fig:ferresdgrm} Ferres diagrams for the constraints on
  the edge partitions for $k = 2$ (left) and $k = 3$ in the
  Read-Rezayi series with $r = 2$ (right). The left Ferres diagram
  illustrates how an edge parition is determined for $k = 2$. In the
  example, the edge partition can be bipartited into
  $\eta^{(1)}=\{7,6,5,5,2\}$ and an $\eta^{(2)}$ to be determined. The
  subpartition represented by the white cells (including the grey
  cells) is $\eta^{(1)}$ and partition represented by the grey is a
  partition obtained by shifting $\eta^{(1)}$ upward by 1 and leftward
  by $r = 2$. As discussed in the text, an allowed $\eta^{(2)}$ should
  be dominated by $\eta^{(1)}$ but dominate the grey-cell
  partition. The right Ferres diagram illustrates the case for $k=3$,
  assuming that $\eta^{(2)}$ is already selected based on the left
  Ferres diagram, e.g., $\eta^{(2)} = \{4,3,3\}$ (grey and dark grey
  cells) for the same $\eta^{(1)} = \{7,6,5,5,2\}$ (white, grey, and
  dark grey cells). $\eta^{(3)}$ should be dominated by $\eta^{(2)}$,
  but dominate the partition represented by dark grey cells, obtained
  by shifting $\eta^{(2)}$ upward by 1 and leftward by $r = 2$.  }
\end{figure}

To discuss $(k,r)$ admissible states with $k > 1$ is more complicated,
as $\eta$ cannot be a simple partition. To systemically classify these
admissible edge states, it is necessary to separate a partition into
$k$ sequences.  By doing so we get a set of $\eta^{(1)}$, ...
,$\eta^{(k)}$, where $\eta^{(i)}$ = $\left\{\eta_i, \eta_{i+k},
\eta_{i+2k}, \cdots, \eta _{n-k+i} \right\}$. The $(k,r)$ admissible
condition of $\lambda$ implies that
\begin{equation}
\eta^{(i \bmod k)}_{\lceil i/k \rceil}-\eta^{(i \bmod k)}_{\lceil
  i/k \rceil + 1} = \eta_{i}-\eta_{i+k} = (\lambda_{i} -
\lambda_{i+k}) - r \geq 0
\label{edgecondition}
\end{equation}
where $\lceil x \rceil$ is the smallest integer greater than or equal
to $x$.  In other words, $\eta^{(i)}$ is a partition of
$\left|\eta^{(i)}\right|$. But unlike the $k=1$ case, not every
partition of a given angular momentum is qualified as $\eta^{i}$.  If
$k \neq 1$ we have two conditions that exist among $\eta^{(i)}_j$s.
The first condition is that $\eta^{(i)}_j \geq \eta^{(i+1)}_j$. This
condition implies that $|\eta^{(i)}| \geq |\eta^{(i+1)}|$, which is
weaker.  So, $ \{|\eta^{(1)}|, |\eta^{(2)}|, \cdots, |\eta^{(k)}|\}$
is itself a partition of $\Delta M$.  The second condition is that
$\eta^{(i)}_{j} < \eta^{(i+1)}_{j+1} + r$ for every $1 \le i \le k$
and $j \ge 1$.  Without such conditions, two different $\eta$ can be
assign to the same state.  For example, the additions of two sets of
partitions $\eta^{(1)} = \{2\}$, $\eta^{(2)} = \{1\}$ and $\xi^{(1)} =
\{1\}$, $\xi^{(2)} = \{2\}$ generate the same state for the Moore-Read
ground state with a densest root configuration
$\{2,0,2,\cdots\}$. Careful examination leads to the conclusion that
the above two conditions are sufficient to generate a unique
representation.  These conditions are easily depicted by a Ferres
diagram in Fig.~\ref{fig:ferresdgrm}.  The counting of $\eta$ is not
as straightforward as in the $k = 1$ case, but can be done by
partitioning the angular momentum of the edge excitation into $k$
ordered partitions as discussed here. The counting of $(k,2)$ and
$(k,3)$ admissible partitions are also considered in Ref.
\onlinecite{bernevig2008properties} to compute the specific heat of
FQH states.

\section{Derivation of Eq.~(\ref{LAEDGEEQ})}
\label{app:LaEdgeEq}
To verify the relation $J_{\Omega + \eta}^{\alpha} = \JE \JO$ where
$\alpha = -\frac{2}{r-1}$ and $\beta = \frac{2}{r+1}$ (i.e. $k=1$), we
start with the Calogero-Sutherland Hamiltonian $H_{CS} = \sum
_iz_iD_i^{L,-1} z_iD_i^{L,r}$, where $D_i^{L,r}=\ri-r\sum _{j (\neq
  i)}\frac{1}{\zi-\zj}$ and, importantly, $D_{i}^{L,r} \JO =
0$,~\cite{bernevig08}, and apply it to $\JE J_{\Omega}^{\alpha} $.
\begin{multline}
\sum_{i} \zi D_{i}^{L,-1} \zi D_{i}^{L,r} \JE \JO = \sum_{i} \zi
D_{i}^{L,-1} \zi \JO \ri \JE \\ = \sum_{i}\left[\zi \JO \ri \JE
  \right. + \zi^{2} \JO \ri^{2} \JE \\ \left. + \zi^{2} \ri \JO \ri
  \JE + \zi^{2} \JO \sum_{j(\neq i)} \frac{\ri \JE }{\zi - \zj}
  \right] \\ = \JO \left( \sum_{i} \left[ \zi \ri \JE + \zi^{2}
  \ri^{2} \JE \right]\right. \\ \left. + (1 + r) \sum_{i \neq j}
\frac{\zi^{2} \ri \JE }{\zi - \zj} \right)\\ = \JO \left( \sum_{i}
(\zi \ri)^{2} + (1+r)\sum_{i \neq j} \frac{\zi^{2} \ri }{\zi-\zj}
\right) \JE.
\end{multline}
The first and the third equalities come from the relation $
D_{i}^{L,r} \JO = 0$. But
\begin{multline}
2 \sum_{i \neq j} \frac{\zi^{2} \ri}{\zi - \zj} - \sum_{i \neq j} \ri
\zi = \sum_{i \neq j} \frac{\left(\zi^{2}+ \zi \zj\right)\ri}{\zi-\zj}
\\ = \sum_{i < j} \frac{\zi + \zj}{\zi-\zj} (\zi \ri - \zj \rj).
\end{multline}
Hence, 
\begin{multline}
\sum_{i \neq j} \frac{\zi^{2} \ri }{\zi - \zj} = \frac{1}{2} \sum_{i <
  j} \frac{\zi + \zj}{\zi-z{j}} (\zi \ri - \zj \rj) \\+\frac{1}{2}
\sum_{i \neq j} \zi \ri.
\end{multline}
In order to be the solution of the equation, $\JE$ is the solution of the
\begin{multline}
\sum_{i} \left(\zi \ri\right)^{2} + \frac{1+r}{2} \sum_{i < j}
\frac{\zi + \zj}{\zi-\zj}(\zi \ri - \zj \rj) \\ + \frac{1+r}{2}
\sum_{i \neq j} \zi \ri.
\end{multline}
It is the Calogero-Sutherland Hamiltonian with a constant offset (note
that $\sum_i \zi \ri$ is nothing but the total angular momentum) and a
positive parameter $ \frac{2}{1+r}$. By choosing $\beta =
\frac{2}{1+r}$ and following the conventional normalization of the
Jack, we complete the proof of Eq.~(\ref{LAEDGEEQ}).

\section{Proof of $H_{sq}(\nu)H_{sq}(\omega) \subset H_{sq}(\nu + \omega)$}
\label{app:proof}
In this appendix we examine the multiplication of two monomials
labeled by $\lambda \in H_{\text{sq}}(\nu)$ and $\mu \in
H_{\text{sq}}(\omega)$. We will show that $m_{\lambda}m_{\mu}$ is in
$H_{\text{sq}}(\nu +\omega)$, i.e., the Hilbert space of monomials
labeled by partitions that can be squeezed from $\nu +\omega$. The
addition of two partitions is understood as element-wise addition. For
example, $\{9,5,2\}+\{4,2\} = \{13,7,2\}$. Obviously, the
multiplication of two monomials is not necessarily a single symmetric
monomial but a sum of them, or a generic symmetric polynomial.

First, we need to examine what kind of symmetric monomial is included
in this generic symmetric polynomial.  Consider the multiplication of
two monomials $m_{\lambda}$ and $m_{\mu}$.
\begin{equation}
m_{\lambda}m_{\mu} =\underset{\lambda,\mu}{\text{Sym }} z_1^{\lambda
  _1+\mu _1}\cdots z_n^{\lambda _n+\mu _n},
\end{equation}
which can then be rewritten as an expansion of symmetric
monomials. Consider an arbitrary symmetric monomial in the expansion.
The key step is to show that the corresponding partition is dominated
by $\lambda +\mu$. To show this, let us fix $\lambda$ and permute
$\mu$. The permutation will generate terms belonging to other
symmetric monomials, but they are related to $m_{\lambda +\mu} $ by
squeezing. After a permutation of $\lambda _i$ and $\lambda _j$ for
example, we obtain a term included in a symmetric monomial
$\left\{\cdots,\lambda_i+\mu _j,\cdots,\lambda _j+\mu _i\right\} $. If
we assume that $i > j$, the permutation brings the momenta of the
corresponding pair of particles closer and the new partition is
dominated by the original one; this is what we call
squeezing. Starting from the monomial $m_{\lambda +\mu}$, we can show
that all terms in $m_{\lambda}m_{\mu}$ can be obtained by squeezing
and, therefore, their corresponding partitions are dominated by
$\lambda +\mu$. Similarly, one can show that $\lambda + \mu$ is
dominated by $\nu +\mu$, which is then dominated by $\nu +\omega$. If
we define the multiplication of two Hilbert spaces $H_{\text{sq}}(\nu
)$ and $H_{\text{sq}}(\omega)$ as
\begin{equation}
 \left\{m: \langle m \vert m_{\lambda}m_{\mu} \rangle \neq 0 , \lambda
 \in H_{\text{sq}}(\nu ),\mu \in H_{\text{sq}}(\omega)\right\},
\end{equation}
we can say that the multiplication of two Hilbert spaces squeezed from
two separate partitions $\nu$ and $\omega$ is included in the Hilbert
space squeezed from the sum of two partitions, i.e.,
\begin{equation}
H_{sq}(\nu)H_{sq}(\omega) \subset H_{sq}(\nu + \omega).
\end{equation}


\begin{thebibliography}{38}
\expandafter\ifx\csname natexlab\endcsname\relax\def\natexlab#1{#1}\fi
\expandafter\ifx\csname bibnamefont\endcsname\relax
  \def\bibnamefont#1{#1}\fi
\expandafter\ifx\csname bibfnamefont\endcsname\relax
  \def\bibfnamefont#1{#1}\fi
\expandafter\ifx\csname citenamefont\endcsname\relax
  \def\citenamefont#1{#1}\fi
\expandafter\ifx\csname url\endcsname\relax
  \def\url#1{\texttt{#1}}\fi
\expandafter\ifx\csname urlprefix\endcsname\relax\def\urlprefix{URL }\fi
\providecommand{\bibinfo}[2]{#2}
\providecommand{\eprint}[2][]{\url{#2}}

\bibitem[{\citenamefont{Laughlin}(1983)}]{laughlin83}
\bibinfo{author}{\bibfnamefont{R.~B.} \bibnamefont{Laughlin}},
  \bibinfo{journal}{Phys. Rev. Lett.} \textbf{\bibinfo{volume}{50}},
  \bibinfo{pages}{1395} (\bibinfo{year}{1983}).

\bibitem[{\citenamefont{Jain}(2007)}]{jainbook}
\bibinfo{author}{\bibfnamefont{J.~K.} \bibnamefont{Jain}},
  \emph{\bibinfo{title}{Composite Fermions}} (\bibinfo{publisher}{Cambridge
  University Press}, \bibinfo{address}{New York}, \bibinfo{year}{2007}), ISBN
  \bibinfo{isbn}{978-0-521-86232-5}.

\bibitem[{\citenamefont{Bernevig and Haldane}(2008{\natexlab{a}})}]{bernevig07}
\bibinfo{author}{\bibfnamefont{B.~A.} \bibnamefont{Bernevig}} \bibnamefont{and}
  \bibinfo{author}{\bibfnamefont{F.~D.~M.} \bibnamefont{Haldane}},
  \bibinfo{journal}{Phys. Rev. Lett.} \textbf{\bibinfo{volume}{100}},
  \bibinfo{pages}{246802} (\bibinfo{year}{2008}{\natexlab{a}}).

\bibitem[{\citenamefont{Chang}(2003)}]{chang03}
\bibinfo{author}{\bibfnamefont{A.~M.} \bibnamefont{Chang}},
  \bibinfo{journal}{Rev. Mod. Phys.} \textbf{\bibinfo{volume}{75}},
  \bibinfo{pages}{1449} (\bibinfo{year}{2003}).

\bibitem[{\citenamefont{Radu et~al.}(2008)\citenamefont{Radu, Miller, Marcus,
  Kastner, Pfeiffer, and West}}]{marcus08}
\bibinfo{author}{\bibfnamefont{I.~P.} \bibnamefont{Radu}},
  \bibinfo{author}{\bibfnamefont{J.~B.} \bibnamefont{Miller}},
  \bibinfo{author}{\bibfnamefont{C.~M.} \bibnamefont{Marcus}},
  \bibinfo{author}{\bibfnamefont{M.~A.} \bibnamefont{Kastner}},
  \bibinfo{author}{\bibfnamefont{L.~N.} \bibnamefont{Pfeiffer}},
  \bibnamefont{and} \bibinfo{author}{\bibfnamefont{K.~W.} \bibnamefont{West}},
  \bibinfo{journal}{Science} \textbf{\bibinfo{volume}{320}},
  \bibinfo{pages}{899} (\bibinfo{year}{2008}).

\bibitem[{\citenamefont{Bid et~al.}(2010)\citenamefont{Bid, Ofek, Inoue,
  Heiblum, Kane et~al.}}]{heiblem10}
\bibinfo{author}{\bibfnamefont{A.}~\bibnamefont{Bid}},
  \bibinfo{author}{\bibfnamefont{N.}~\bibnamefont{Ofek}},
  \bibinfo{author}{\bibfnamefont{H.}~\bibnamefont{Inoue}},
  \bibinfo{author}{\bibfnamefont{M.}~\bibnamefont{Heiblum}},
  \bibinfo{author}{\bibfnamefont{C.}~\bibnamefont{Kane}}, \bibnamefont{et~al.},
  \bibinfo{journal}{Nature} \textbf{\bibinfo{volume}{466}},
  \bibinfo{pages}{585} (\bibinfo{year}{2010}), \eprint{1005.5724}.

\bibitem[{\citenamefont{Willett et~al.}(2009)\citenamefont{Willett, Pfeiffer,
  and West}}]{willett09}
\bibinfo{author}{\bibfnamefont{R.~L.} \bibnamefont{Willett}},
  \bibinfo{author}{\bibfnamefont{L.~N.} \bibnamefont{Pfeiffer}},
  \bibnamefont{and} \bibinfo{author}{\bibfnamefont{K.~W.} \bibnamefont{West}},
  \bibinfo{journal}{Proceedings of the National Academy of Sciences}
  \textbf{\bibinfo{volume}{106}}, \bibinfo{pages}{8853} (\bibinfo{year}{2009}).

\bibitem[{\citenamefont{Willett et~al.}(2010)\citenamefont{Willett, Pfeiffer,
  and West}}]{willett10}
\bibinfo{author}{\bibfnamefont{R.~L.} \bibnamefont{Willett}},
  \bibinfo{author}{\bibfnamefont{L.~N.} \bibnamefont{Pfeiffer}},
  \bibnamefont{and} \bibinfo{author}{\bibfnamefont{K.~W.} \bibnamefont{West}},
  \bibinfo{journal}{Phys. Rev. B} \textbf{\bibinfo{volume}{82}},
  \bibinfo{pages}{205301} (\bibinfo{year}{2010}).

\bibitem[{\citenamefont{Willett et~al.}({\natexlab{a}})\citenamefont{Willett,
  Pfeiffer, and West}}]{willett12}
\bibinfo{author}{\bibfnamefont{R.}~\bibnamefont{Willett}},
  \bibinfo{author}{\bibfnamefont{L.}~\bibnamefont{Pfeiffer}}, \bibnamefont{and}
  \bibinfo{author}{\bibfnamefont{K.}~\bibnamefont{West}},
  \bibinfo{note}{arXiv:1204.1993 (unpublished)}.

\bibitem[{\citenamefont{Li and Haldane}(2008)}]{li08}
\bibinfo{author}{\bibfnamefont{H.}~\bibnamefont{Li}} \bibnamefont{and}
  \bibinfo{author}{\bibfnamefont{F.~D.~M.} \bibnamefont{Haldane}},
  \bibinfo{journal}{Phys. Rev. Lett.} \textbf{\bibinfo{volume}{101}},
  \bibinfo{pages}{010504} (\bibinfo{year}{2008}).

\bibitem[{\citenamefont{Dubail et~al.}(2012{\natexlab{a}})\citenamefont{Dubail,
  Read, and Rezayi}}]{dubail12a}
\bibinfo{author}{\bibfnamefont{J.}~\bibnamefont{Dubail}},
  \bibinfo{author}{\bibfnamefont{N.}~\bibnamefont{Read}}, \bibnamefont{and}
  \bibinfo{author}{\bibfnamefont{E.~H.} \bibnamefont{Rezayi}},
  \bibinfo{journal}{Phys. Rev. B} \textbf{\bibinfo{volume}{85}},
  \bibinfo{pages}{115321} (\bibinfo{year}{2012}{\natexlab{a}}).

\bibitem[{\citenamefont{Sterdyniak et~al.}(2012)\citenamefont{Sterdyniak,
  Chandran, Regnault, Bernevig, and Bonderson}}]{bernevig12}
\bibinfo{author}{\bibfnamefont{A.}~\bibnamefont{Sterdyniak}},
  \bibinfo{author}{\bibfnamefont{A.}~\bibnamefont{Chandran}},
  \bibinfo{author}{\bibfnamefont{N.}~\bibnamefont{Regnault}},
  \bibinfo{author}{\bibfnamefont{B.~A.} \bibnamefont{Bernevig}},
  \bibnamefont{and}
  \bibinfo{author}{\bibfnamefont{P.}~\bibnamefont{Bonderson}},
  \bibinfo{journal}{Phys. Rev. B} \textbf{\bibinfo{volume}{85}},
  \bibinfo{pages}{125308} (\bibinfo{year}{2012}).

\bibitem[{\citenamefont{Rodr{\'i}guez et~al.}(2012)\citenamefont{Rodr{\'i}guez,
  Simon, and Slingerland}}]{simon12}
\bibinfo{author}{\bibfnamefont{I.~D.} \bibnamefont{Rodr{\'i}guez}},
  \bibinfo{author}{\bibfnamefont{S.~H.} \bibnamefont{Simon}}, \bibnamefont{and}
  \bibinfo{author}{\bibfnamefont{J.~K.} \bibnamefont{Slingerland}},
  \bibinfo{journal}{Phys. Rev. Lett.} \textbf{\bibinfo{volume}{108}},
  \bibinfo{pages}{256806} (\bibinfo{year}{2012}).

\bibitem[{\citenamefont{Dubail et~al.}(2012{\natexlab{b}})\citenamefont{Dubail,
  Read, and Rezayi}}]{dubail12}
\bibinfo{author}{\bibfnamefont{J.}~\bibnamefont{Dubail}},
  \bibinfo{author}{\bibfnamefont{N.}~\bibnamefont{Read}}, \bibnamefont{and}
  \bibinfo{author}{\bibfnamefont{E.~H.} \bibnamefont{Rezayi}},
  \bibinfo{journal}{Phys. Rev. B} \textbf{\bibinfo{volume}{86}},
  \bibinfo{pages}{245310} (\bibinfo{year}{2012}{\natexlab{b}}).

\bibitem[{\citenamefont{Kitaev and Preskill}(2006)}]{kitaev06}
\bibinfo{author}{\bibfnamefont{A.}~\bibnamefont{Kitaev}} \bibnamefont{and}
  \bibinfo{author}{\bibfnamefont{J.}~\bibnamefont{Preskill}},
  \bibinfo{journal}{Phys. Rev. Lett.} \textbf{\bibinfo{volume}{96}},
  \bibinfo{pages}{110404} (\bibinfo{year}{2006}).

\bibitem[{\citenamefont{Wen}(1992)}]{wen1992theory}
\bibinfo{author}{\bibfnamefont{X.-G.} \bibnamefont{Wen}},
  \bibinfo{journal}{Int. J. Mod. Phys. B} \textbf{\bibinfo{volume}{6}},
  \bibinfo{pages}{1711} (\bibinfo{year}{1992}).

\bibitem[{\citenamefont{{Milovanovi\ifmmode \acute{c}\else {\'c}\fi{}} and
  Read}(1996)}]{read96}
\bibinfo{author}{\bibfnamefont{M.}~\bibnamefont{{Milovanovi\ifmmode
  \acute{c}\else {\'c}\fi{}}}} \bibnamefont{and}
  \bibinfo{author}{\bibfnamefont{N.}~\bibnamefont{Read}},
  \bibinfo{journal}{Phys. Rev. B} \textbf{\bibinfo{volume}{53}},
  \bibinfo{pages}{13559} (\bibinfo{year}{1996}).

\bibitem[{\citenamefont{Wan et~al.}(2003)\citenamefont{Wan, Rezayi, and
  Yang}}]{wan03}
\bibinfo{author}{\bibfnamefont{X.}~\bibnamefont{Wan}},
  \bibinfo{author}{\bibfnamefont{E.~H.} \bibnamefont{Rezayi}},
  \bibnamefont{and} \bibinfo{author}{\bibfnamefont{K.}~\bibnamefont{Yang}},
  \bibinfo{journal}{Phys. Rev. B} \textbf{\bibinfo{volume}{68}},
  \bibinfo{pages}{125307} (\bibinfo{year}{2003}).

\bibitem[{\citenamefont{Wan et~al.}(2008)\citenamefont{Wan, Hu, Rezayi, and
  Yang}}]{wan08}
\bibinfo{author}{\bibfnamefont{X.}~\bibnamefont{Wan}},
  \bibinfo{author}{\bibfnamefont{Z.-X.} \bibnamefont{Hu}},
  \bibinfo{author}{\bibfnamefont{E.~H.} \bibnamefont{Rezayi}},
  \bibnamefont{and} \bibinfo{author}{\bibfnamefont{K.}~\bibnamefont{Yang}},
  \bibinfo{journal}{Phys. Rev. B} \textbf{\bibinfo{volume}{77}},
  \bibinfo{pages}{165316} (\bibinfo{year}{2008}).

\bibitem[{\citenamefont{Jolad et~al.}(2010)\citenamefont{Jolad, Sen, and
  Jain}}]{jain10}
\bibinfo{author}{\bibfnamefont{S.}~\bibnamefont{Jolad}},
  \bibinfo{author}{\bibfnamefont{D.}~\bibnamefont{Sen}}, \bibnamefont{and}
  \bibinfo{author}{\bibfnamefont{J.~K.} \bibnamefont{Jain}},
  \bibinfo{journal}{Phys. Rev. B} \textbf{\bibinfo{volume}{82}},
  \bibinfo{pages}{075315} (\bibinfo{year}{2010}).

\bibitem[{\citenamefont{Bernevig and Haldane}(2008{\natexlab{b}})}]{bernevig08}
\bibinfo{author}{\bibfnamefont{B.~A.} \bibnamefont{Bernevig}} \bibnamefont{and}
  \bibinfo{author}{\bibfnamefont{F.~D.~M.} \bibnamefont{Haldane}},
  \bibinfo{journal}{Phys. Rev. Lett.} \textbf{\bibinfo{volume}{100}},
  \bibinfo{pages}{246802} (\bibinfo{year}{2008}{\natexlab{b}}).

\bibitem[{\citenamefont{Bernevig and
  Haldane}(2008{\natexlab{c}})}]{bernevig2008properties}
\bibinfo{author}{\bibfnamefont{B.~A.} \bibnamefont{Bernevig}} \bibnamefont{and}
  \bibinfo{author}{\bibfnamefont{F.~D.~M.} \bibnamefont{Haldane}},
  \bibinfo{journal}{Phys. Rev. Lett.} \textbf{\bibinfo{volume}{101}},
  \bibinfo{pages}{246806} (\bibinfo{year}{2008}{\natexlab{c}}).

\bibitem[{\citenamefont{Bernevig and Regnault}(2009)}]{bernevig09}
\bibinfo{author}{\bibfnamefont{B.~A.} \bibnamefont{Bernevig}} \bibnamefont{and}
  \bibinfo{author}{\bibfnamefont{N.}~\bibnamefont{Regnault}},
  \bibinfo{journal}{Phys. Rev. Lett.} \textbf{\bibinfo{volume}{103}},
  \bibinfo{pages}{206801} (\bibinfo{year}{2009}).

\bibitem[{\citenamefont{Feigin et~al.}(2002)\citenamefont{Feigin, Jimbo, Miwa,
  and Mukhin}}]{feigin2002differential}
\bibinfo{author}{\bibfnamefont{B.}~\bibnamefont{Feigin}},
  \bibinfo{author}{\bibfnamefont{M.}~\bibnamefont{Jimbo}},
  \bibinfo{author}{\bibfnamefont{T.}~\bibnamefont{Miwa}}, \bibnamefont{and}
  \bibinfo{author}{\bibfnamefont{E.}~\bibnamefont{Mukhin}},
  \bibinfo{journal}{Int. Math. Res. Not.} \textbf{\bibinfo{volume}{2002}},
  \bibinfo{pages}{1223} (\bibinfo{year}{2002}).

\bibitem[{\citenamefont{Feigin et~al.}(2003)\citenamefont{Feigin, Jimbo, Miwa,
  and Mukhin}}]{feigin2003symmetric}
\bibinfo{author}{\bibfnamefont{B.}~\bibnamefont{Feigin}},
  \bibinfo{author}{\bibfnamefont{M.}~\bibnamefont{Jimbo}},
  \bibinfo{author}{\bibfnamefont{T.}~\bibnamefont{Miwa}}, \bibnamefont{and}
  \bibinfo{author}{\bibfnamefont{E.}~\bibnamefont{Mukhin}},
  \bibinfo{journal}{Int. Math. Res. Not.} \textbf{\bibinfo{volume}{2003}},
  \bibinfo{pages}{1015} (\bibinfo{year}{2003}).

\bibitem[{\citenamefont{Estienne and Santachiara}(2009)}]{estienne2009relating}
\bibinfo{author}{\bibfnamefont{B.}~\bibnamefont{Estienne}} \bibnamefont{and}
  \bibinfo{author}{\bibfnamefont{R.}~\bibnamefont{Santachiara}},
  \bibinfo{journal}{J. Phys. A: Math. Theor.} \textbf{\bibinfo{volume}{42}},
  \bibinfo{pages}{445209} (\bibinfo{year}{2009}).

\bibitem[{\citenamefont{Yang et~al.}(2012)\citenamefont{Yang, Hu, {Papi\ifmmode
  \acute{c}\else {\'c}\fi{}}, and Haldane}}]{yang12}
\bibinfo{author}{\bibfnamefont{B.}~\bibnamefont{Yang}},
  \bibinfo{author}{\bibfnamefont{Z.-X.} \bibnamefont{Hu}},
  \bibinfo{author}{\bibfnamefont{Z.}~\bibnamefont{{Papi\ifmmode \acute{c}\else
  {\'c}\fi{}}}}, \bibnamefont{and} \bibinfo{author}{\bibfnamefont{F.~D.~M.}
  \bibnamefont{Haldane}}, \bibinfo{journal}{Phys. Rev. Lett.}
  \textbf{\bibinfo{volume}{108}}, \bibinfo{pages}{256807}
  (\bibinfo{year}{2012}).

\bibitem[{\citenamefont{Bernevig and
  Haldane}(2008{\natexlab{d}})}]{bernevig2008generalized}
\bibinfo{author}{\bibfnamefont{B.~A.} \bibnamefont{Bernevig}} \bibnamefont{and}
  \bibinfo{author}{\bibfnamefont{F.~D.~M.} \bibnamefont{Haldane}},
  \bibinfo{journal}{Phys. Rev. B} \textbf{\bibinfo{volume}{77}},
  \bibinfo{pages}{184502} (\bibinfo{year}{2008}{\natexlab{d}}).

\bibitem[{\citenamefont{Dumitriu et~al.}(2007)\citenamefont{Dumitriu, Edelman,
  and Shuman}}]{dumitriu2007mops}
\bibinfo{author}{\bibfnamefont{I.}~\bibnamefont{Dumitriu}},
  \bibinfo{author}{\bibfnamefont{A.}~\bibnamefont{Edelman}}, \bibnamefont{and}
  \bibinfo{author}{\bibfnamefont{G.}~\bibnamefont{Shuman}},
  \bibinfo{journal}{J. Sym. Comp.} \textbf{\bibinfo{volume}{42}},
  \bibinfo{pages}{587} (\bibinfo{year}{2007}).

\bibitem[{\citenamefont{{Thomale} et~al.}(2011)\citenamefont{{Thomale},
  {Estienne}, {Regnault}, and {Bernevig}}}]{thomale10decompose}
\bibinfo{author}{\bibfnamefont{R.}~\bibnamefont{{Thomale}}},
  \bibinfo{author}{\bibfnamefont{B.}~\bibnamefont{{Estienne}}},
  \bibinfo{author}{\bibfnamefont{N.}~\bibnamefont{{Regnault}}},
  \bibnamefont{and} \bibinfo{author}{\bibfnamefont{B.~A.}
  \bibnamefont{{Bernevig}}}, \bibinfo{journal}{Phys. Rev. B}
  \textbf{\bibinfo{volume}{84}}, \bibinfo{pages}{045127}
  (\bibinfo{year}{2011}).

\bibitem[{\citenamefont{Bernevig and Haldane}(2009)}]{bernevig2009clustering}
\bibinfo{author}{\bibfnamefont{B.~A.} \bibnamefont{Bernevig}} \bibnamefont{and}
  \bibinfo{author}{\bibfnamefont{F.~D.~M.} \bibnamefont{Haldane}},
  \bibinfo{journal}{Phys. Rev. Lett.} \textbf{\bibinfo{volume}{102}},
  \bibinfo{pages}{66802} (\bibinfo{year}{2009}).

\bibitem[{\citenamefont{Bernevig et~al.}(2009)\citenamefont{Bernevig, Gurarie,
  and Simon}}]{bernevig2009central}
\bibinfo{author}{\bibfnamefont{B.~A.} \bibnamefont{Bernevig}},
  \bibinfo{author}{\bibfnamefont{V.}~\bibnamefont{Gurarie}}, \bibnamefont{and}
  \bibinfo{author}{\bibfnamefont{S.~H.} \bibnamefont{Simon}},
  \bibinfo{journal}{J. Phys. A: Math. Theor.} \textbf{\bibinfo{volume}{42}},
  \bibinfo{pages}{245206} (\bibinfo{year}{2009}).

\bibitem[{\citenamefont{Estienne et~al.}(2010)\citenamefont{Estienne, Bernevig,
  and Santachiara}}]{estienne2010electron}
\bibinfo{author}{\bibfnamefont{B.}~\bibnamefont{Estienne}},
  \bibinfo{author}{\bibfnamefont{B.~A.} \bibnamefont{Bernevig}},
  \bibnamefont{and}
  \bibinfo{author}{\bibfnamefont{R.}~\bibnamefont{Santachiara}},
  \bibinfo{journal}{Phys. Rev. B} \textbf{\bibinfo{volume}{82}},
  \bibinfo{pages}{205307} (\bibinfo{year}{2010}).

\bibitem[{\citenamefont{Hu et~al.}(2009)\citenamefont{Hu, Rezayi, Wan, and
  Yang}}]{hu2009edge}
\bibinfo{author}{\bibfnamefont{Z.-X.} \bibnamefont{Hu}},
  \bibinfo{author}{\bibfnamefont{E.~H.} \bibnamefont{Rezayi}},
  \bibinfo{author}{\bibfnamefont{X.}~\bibnamefont{Wan}}, \bibnamefont{and}
  \bibinfo{author}{\bibfnamefont{K.}~\bibnamefont{Yang}},
  \bibinfo{journal}{Phys. Rev. B} \textbf{\bibinfo{volume}{80}},
  \bibinfo{pages}{235330} (\bibinfo{year}{2009}).

\bibitem[{\citenamefont{Willett et~al.}({\natexlab{b}})\citenamefont{Willett,
  Pfeiffer, West, and Manfra}}]{willett13}
\bibinfo{author}{\bibfnamefont{R.}~\bibnamefont{Willett}},
  \bibinfo{author}{\bibfnamefont{L.}~\bibnamefont{Pfeiffer}},
  \bibinfo{author}{\bibfnamefont{K.}~\bibnamefont{West}}, \bibnamefont{and}
  \bibinfo{author}{\bibfnamefont{M.}~\bibnamefont{Manfra}},
  \bibinfo{note}{arXiv:1301.2594 (unpublished)}.

\bibitem[{\citenamefont{Xia et~al.}(2004)\citenamefont{Xia, Pan, Vicente,
  Adams, Sullivan, Stormer, Tsui, Pfeiffer, Baldwin, and West}}]{xia04}
\bibinfo{author}{\bibfnamefont{J.~S.} \bibnamefont{Xia}},
  \bibinfo{author}{\bibfnamefont{W.}~\bibnamefont{Pan}},
  \bibinfo{author}{\bibfnamefont{C.~L.} \bibnamefont{Vicente}},
  \bibinfo{author}{\bibfnamefont{E.~D.} \bibnamefont{Adams}},
  \bibinfo{author}{\bibfnamefont{N.~S.} \bibnamefont{Sullivan}},
  \bibinfo{author}{\bibfnamefont{H.~L.} \bibnamefont{Stormer}},
  \bibinfo{author}{\bibfnamefont{D.}~\bibnamefont{Tsui}},
  \bibinfo{author}{\bibfnamefont{L.~N.} \bibnamefont{Pfeiffer}},
  \bibinfo{author}{\bibfnamefont{K.~W.} \bibnamefont{Baldwin}},
  \bibnamefont{and} \bibinfo{author}{\bibfnamefont{K.~W.} \bibnamefont{West}},
  \bibinfo{journal}{Phys. Rev. Lett} \textbf{\bibinfo{volume}{93}},
  \bibinfo{pages}{176809} (\bibinfo{year}{2004}).

\bibitem[{\citenamefont{Kumar et~al.}(2010)\citenamefont{Kumar, Csathy, Manfra,
  Pfeiffer, and West}}]{kumar10}
\bibinfo{author}{\bibfnamefont{A.}~\bibnamefont{Kumar}},
  \bibinfo{author}{\bibfnamefont{G.~A.} \bibnamefont{Csathy}},
  \bibinfo{author}{\bibfnamefont{M.~J.} \bibnamefont{Manfra}},
  \bibinfo{author}{\bibfnamefont{L.~N.} \bibnamefont{Pfeiffer}},
  \bibnamefont{and} \bibinfo{author}{\bibfnamefont{K.~W.} \bibnamefont{West}},
  \bibinfo{journal}{Phys. Rev. Lett} \textbf{\bibinfo{volume}{105}},
  \bibinfo{pages}{246808} (\bibinfo{year}{2010}).

\bibitem[{\citenamefont{Zhang et~al.}(2012)\citenamefont{Zhang, Huan, Xia,
  Sullivan, Pan, Baldwin, West, Pfeiffer, and Tsui}}]{Zhangchi2012}
\bibinfo{author}{\bibfnamefont{C.}~\bibnamefont{Zhang}},
  \bibinfo{author}{\bibfnamefont{C.}~\bibnamefont{Huan}},
  \bibinfo{author}{\bibfnamefont{J.~S.} \bibnamefont{Xia}},
  \bibinfo{author}{\bibfnamefont{N.~S.} \bibnamefont{Sullivan}},
  \bibinfo{author}{\bibfnamefont{W.}~\bibnamefont{Pan}},
  \bibinfo{author}{\bibfnamefont{K.~W.} \bibnamefont{Baldwin}},
  \bibinfo{author}{\bibfnamefont{K.~W.} \bibnamefont{West}},
  \bibinfo{author}{\bibfnamefont{L.~N.} \bibnamefont{Pfeiffer}},
  \bibnamefont{and} \bibinfo{author}{\bibfnamefont{D.~C.} \bibnamefont{Tsui}},
  \bibinfo{journal}{Phys. Rev. B} \textbf{\bibinfo{volume}{85}},
  \bibinfo{pages}{241302} (\bibinfo{year}{2012}).

\end{thebibliography}

\end{document}